# Genome hybridization: A universal way for the origin and diversification of organelles as well as the origin and speciation of eukaryotes

Qing-lin Dong* & Xiang-ying Xing

Department of Bioengineering, Hebei University of Technology, Tianjin, 300130, China

*Corresponding author: Qing-lin Dong (qldong@hebut.edu.cn)

## Abstract

The origin of organelles (mitochondrion, chloroplast and nucleus) remains enigmatic. The endosymbiotic hypothesis that chloroplasts, mitochondria and nuclei descend from the endosymbiotic cyanobacterium, bacterium and archaebacterium respectively is dominant yet uncompelling, while our discovery of de novo organelle biogenesis in the cyanobacterium TDX16 that had acquired its green algal host *Haematococcus pluvialis*'s genome disproves this hypothesis. In light of organelle biogenesis in the cyanobacterium TDX16 in combination with the relevant cellular and molecular evidence, we propose genome hybridization hypothesis (GHH) that the origin of organelles and origin of eukaryotes as well as the diversification of organelles and speciation of eukaryotes are unified and achieved by genome hybridization: the endosymbiotic cyanobacteria/bacteria obtain their archaebacterial or eukaryotic hosts' genomes and hybridize with their own ones resulting in expanded genomes containing a mixture of hybrid prokaryotic genes and eukaryotic genes, and thus have to compartmentalize to accommodate different genes for specialized function of photosynthesis (chloroplast), respiration (mitochondrion) and DNA preservation (nucleus), and consequently turn into photosynthetic/heterotrophic eukaryotes. Accordingly, eukaryotes and their organelles are of multiple origins, while the formation of cancer cells is the speciation of eukaryotes as cancer cells are new species of unicellular eukaryotes arising from bacteria. Therefore, GHH provides a theoretical framework unifying evolutionary biology, cancer biology and cell biology and directing the integrated multidisciplinary research.

**Keywords:** Genome, hybridization, origin, diversification, speciation, organelle, eukaryote, cancer cell, bacteria, cyanobacteria, archaebacteria

# 1. Introduction

The origin of DNA-containing organelles (organelles, thereafter), including nucleus, mitochondrion and chloroplast, has been an enigma ever since their discovery with light microscopes. To explain this puzzle numerous hypotheses have been proposed, which are basically categorized into two types: the endosymbiotic hypothesis and the autogenous hypothesis.

## 1.1 The endosymbiotic hypothesis

The endosymbiotic hypothesis proposes that chloroplasts and mitochondria were derived from an endosymbiotic cyanobacterium and an endosymbiotic bacterium respectively. This hypothesis was initially proposed by Mereschkowsky for chloroplast in 1905[1] and Wallin for mitochondrion in 1923[2] based on the similarities between chloroplasts/mitochondria and cyanobacteria/bacteria, restated in the early 1960s accompanying the detection of DNA in chloroplasts[3]and mitochondria [4] with electron microscopes, and popularized in the late 1960s by Margulis (Sagan) [5].

Aside from the endosymbiotic origin of chloroplast, Mereschkowsky also speculated that nucleus evolved from a prokaryote (mycoplasma) engulfed by an amoeboid cell [1], which received less attention as nucleus, unlike chloroplast and mitochondrion, is unique in nature and not analogous to any prokaryotes. Yet with the identification of archaebacteria [6], it is found that the eukaryote genome is a chimera containing both archaebacterial and bacterial genes [7]. As such, it is postulated that nucleus arose from an archaebacterium engulfed by or symbiotic with gram-negative bacteria [8-11].

## 1.2 The autogenous hypothesis

The autogenous hypothesis postulates that mitochondrion and chloroplast originated by compartmentalizing a proto-eukaryote's cytoplasmic DNA with the invaginated cell membrane [12-13] or thylakoid membrane within a cyanobacterial cell [14-15]. Similarly, nucleus arose by enclosing the cellular DNA with the internalized plasma membrane [16] or thylakoid membrane [14-15] and or bacterial vesicle membrane [17].

## 1.3 The endosymbiotic hypothesis is dominant but not compelling

The existence of DNA in chloroplasts and mitochondria and the resemblance of chloroplast/mitochondrial genes and protein synthesis systems (e.g., ribosome) to those of cyanobacteria/bacteria are in favor of the endosymbiotic hypothesis but not the autogenous hypothesis that imply that all genes in eukaryotic cells arose solely by gene duplication. Hence, the endosymbiotic hypothesis for the origin of chloroplast and mitochondrion has become dominant since the late 1970s [18], while the endosymbiotic hypothesis for the origin of nucleus is now prevailing. Nevertheless, the endosymbiotic hypothesis is countered by increasing cellular and molecular evidence and thus not compelling.

## 2.Evidence contradicting the endosymbiotic hypothesis

### 2.1 The organelle envelopes are not derived from the prokaryotic cell envelops

The double-membraned envelopes of mitochondria and chloroplasts are interpreted to derive from the cell envelopes of the endosymbiotic bacterium and cyanobacterium respectively [16,19]. This interpretation is entirely false because the prokaryotic cell envelope consists of cell wall and cytoplasmic membrane [20] that are structurally and functionally different compartments separated by the extracytoplasmic space. The cell wall of cyanobacteria and gram-negative bacteria comprises an outer membrane, an inner peptidoglycan layer and an intervening space [21-22]. If the outer membranes become the corresponding membranes of chloroplast and mitochondrial envelopes, the peptidoglycan layer must be degraded, which is, however, impossible as synthesis inhibition (e.g., by penicillin and other β-lactam antibiotics) or degradation (e.g. by lysozyme) of the peptidoglycan layer results in cell lysis [23]. So, the double-membraned mitochondrial and chloroplast envelopes are not derived form the cell envelopes of the endosymbiotic bacterium and cyanobacterium, which is why many enzymes for synthesizing chloroplast lipids are not of cyanobacterial origin [24]. The double-membraned nuclear envelope derives neither from the cell envelope of the endosymbiotic archaebacterium since its membrane lipids are ether-linked instead of ester-linked as those of bacteria/cyanobacteria and eukaryotes [7], nor from the cell envelope of bacterial host for the same reason as discussed above for chloroplast and mitochondrial envelopes.

### 2.2 The eukaryotic cytoplasmic matrix can not be formed

If the nucleus is derived from an archaebacterium within a bacterium, then, how the eukaryotic cytoplasmic matrix that is compositionally and functionally distinct from the prokaryotic cytoplasm forms and replaces the original bacterial cytoplasm? This is an unanswered yet crucial question as nucleus can not work in the prokaryotic cytoplasm and the eukaryotic cell is not simply the prokaryotic cytoplasm containing a nucleus and other organelles. If the endosymbiotic archaebacterium keeps intact, the eukaryotic cytoplasmic matrix can not be formed, while if the endosymbiotic archaebacterium ruptures for the presumptive merger with the bacterial host, the nucleus can not derive from the archaebacterium.

### 2.3 Only simultaneous origin of organelles is reasonable yet unattainable

The nucleus, mitochondrion and chloroplast are supposed to originate individually in separate symbiotic events. As such the origin order of these organelles has become an issue of debate, particularly the origin sequence of nucleus and mitochondrion. For heterotrophic eukaryotes, it is traditionally assumed that the nucleus came first and later mitochondrion was formed in the nucleus-bearing cell [19], this scenario is argued to be unfeasible due to the insufficient energy supply in the absence of mitochondrion to underpin nucleus development; it is then proposed that mitochondrion predated nucleus in origin and was formed in a prokaryote [25]. Such a scenario is, however, also unattainable, because mitochondria arose putatively from the endosymbiotic bacterium by losing and transferring genes into the nucleus and works only in the eukaryotic cytoplasmic matrix that is quite distinct from the prokaryotic cytoplasm, otherwise it is a bacterium rather than a mitochondrion. Hence, only simultaneous origin of nucleus and mitochondrion seems to be reasonable. Likewise, it is also the case for photosynthetic eukaryotes, in which the ancestral chloroplast, nucleus and mitochondrion were formed simultaneously. Nonetheless, simultaneous origin of these organelles means intact integration of an archaebacterium and a bacterium into a bacterial host cell for the origin of heterotrophic eukaryotes and an additional cyanobacterium into the same bacterial host cell at the same time for the origin photosynthetic eukaryotes, which is mechanically and cell biologically unachievable.

### 2.4 The genomes of mitochondrion, chloroplast and nucleus are not bacterial, cyanobacterial and archaebacterial ones

Different organelles have common DNA sequences termed promiscuous DNA[26], particularly mitochondria and chloroplasts in photosynthetic eukaryotes [27-39], e.g. mitochondrial and chloroplast genomes of *Haematococcus lacustris* are made up of nearly identical repetitive DNA sequences [37], and mitochondria and nuclei in heterotrophic eukaryotes [40-50], where even entire mitochondrial DNAs reside in the nuclei of human [48], rodent [49] and HeLa cell (cancer cell) [43]. Hence the genomes of mitochondrion, chloroplast and nucleus are not bacterial, cyanobacterial and archaebacterial ones.

### 2.5 The proteomes of mitochondrion, chloroplast and nucleus are not bacterial, cyanobacterial and archaebacterial ones

Only a small fraction of the mitochondrial proteins have bacterial homologs, while the rest display no homology to bacterial proteins and thus is not of bacterial origin [51-56]. Also, promiscuous proteins are widespread among organelles. A number of mitochondrial proteins reside in chloroplasts of photosynthetic eukaryotes [57-59] but nuclei of heterotrophic eukaryotes [60-61]. Thus, the proteomes of mitochondrion, chloroplast and nucleus are also not bacterial, cyanobacterial and archaebacterial ones.

### 2.6 Mitochondria present in chloroplasts of photosynthetic eukaryotes but nuclei of heterotrophic eukaryotes

The presence of mitochondria in chloroplasts of photosynthetic eukaryotes [62-67] but nuclei of heterotrophic eukaryotes [68-75] have been frequently observed, yet the reason and mechanism remain unclear. It is most likely that mitochondria are formed in chloroplasts and nuclei, since detachment of mitochondria or mitochondrion-like bodies from chloroplasts in photosynthetic eukaryotes [64,76] but nuclei in heterotrophic eukaryotes [68] have also been detected.

Obviously, the endosymbiotic hypothesis fails to account for the above evidence. The origins of organelles are, in essence, the biogenesis of organelles in prokaryotes and thus are cellular events (a topic of cell biology). Yet, such cellular events or the like and even the intermediate state of cells had not been observed previously. In this circumstance, it is trying to infer the cellular events of organelle origin in evolutionary biology by comparing the similarities between the extant eukaryotes' organelles and prokaryotes in morphology, structure, physiology, genes/genome, proteins/proteome and other molecular data, which is completely speculative and indeed unachievable, because how the extant eukaryotes/organelles descend from the ancestral eukaryotes/organelles, the basic mechanism necessary for inference, is unknown.

In our previous studies, we have found de novo organelle biogenesis in the endosymbiotic cyanobacterium TDX16 that had acquired its green algal host *Haematococcus pluvialis*'s total DNA (genome), resulting in the transition of prokaryote TDX16 into a new species of eukaryote TDX16-DE [77]. This is the first case of organelle biogenesis in prokaryotes observed so far, which provides evidence and an unprecedented reference for inferring the origin of organelles.

## 3. De novo organelle biogenesis in the cyanobacterium TDX16

In brief, organelle biogenesis in the cyanobacterium TDX16 initiated with cytoplasm compartmentalization, followed by de-compartmentalization, DNA hybridization and allocation, and re-compartmentalization, as such only two composite organelles-the primitive chloroplast and primitive nucleus sequestering minor and major fractions of cellular DNA respectively were formed. Thereafter, the eukaryotic cytoplasmic matrix was built up from the matrix extruded from the primitive nucleus; mitochondria were assembled in and segregated from the primitive chloroplast, whereby the primitive nucleus and primitive chloroplast matured into nucleus and chloroplast respectively, leading to the transition of prokaryotic cyanobacterium TDX16 into a new eukaryotic green alga TDX16-DE [77] that is taxonomically assigned as *Chroococcidiorella tianjinensis* [78]. Hence, de novo organelle biogenesis in the cyanobacterium TDX16 is a cellular event for the formation of new organelles (organelle diversification) and new eukaryotic species (speciation), which disproves the endosymbiotic hypothesis and sheds light on the questions that is crucial for inferring the origin of organelles:

(1) Endosymbiosis is prokaryotes' way for bulk acquisition of foreign DNA (i.e. the host cell's total DNA), which is different from conjugation, transformation and transduction [79] that are prokaryotes' ways for piecemeal acquisition of DNA fragments.

(2)Prokaryotes can hybridize the obtained eukaryotic genome with its own one, resulting in loss and retention of some genes as well as synthesis of new hybrid prokaryotic genes and eukaryotic genes, and allocate the retained and hybrid genes selectively into different compartments during compartmentalization.

(3) Chloroplast, nucleus and mitochondrion are formed all at once in a short period of time by enclosing the allocated DNA with the membranes synthesized by fusion, flattening and extension of the small vesicles. Such that, the envelope of these organelles inevitably comprise two unit membranes if the vesicles are bounded by one unit membrane (e.g., vesicles derived from the cytoplasmic membrane and thylakoids), but four unit membranes (e.g., the nuclear envelop of TDX16-DE), if the vesicles are bounded by two unit membranes (e.g., vesicles derived from the primitive chloroplast envelope).
(4)Chloroplast and mitochondrion are both developed from the primitive chloroplast.
(5)The eukaryotic cytoplasmic matrix and nuclear matrix both originate from the matrix of primitive nucleus.
(6)Thylakoids in cyanobacteria are developed initially from the osmiophilic granules assembled on the cytoplasmic membrane (primary thylakoids) and subsequently from the primary thylakoid-derived vesicles or segments (secondary thylakoids).
(7) Thylakoids in chloroplasts are developed originally from the modified osmiophilic granules-the plastoglobuli generated on the transitional secondary thylakoids.
(8) The cristae in mitochondria are formed by fusing, flattening and elongating the single-membrane-bounded vesicles, and thus, like chloroplast thylakoids, are single membrane-bounded sacs, rather than the folds of inner mitochondrial membrane.

## 4. Presentation of the hypothesis

### 4.1 Concept definition: origin/diversification of organelles and origin/speciation of eukaryotes

Conceptually, the origin of organelles (biogenesis of organelles in prokaryotes without the involvement of eukaryotes) and diversification of organelles (biogenesis of organelles in prokaryotes, e.g. TDX16, with the involvement of eukaryotes, e.g., *H. pluvialis*) are two similar yet different cellular events resulting in the origin of eukaryotes (formation of eukaryotes from prokaryotes without the involvement of other eukaryotes) and speciation of eukaryotes (formation of new eukaryotic species, e.g., TDX16-DE, with the involvement of other eukaryotes) respectively.

### 4.2 The origin/diversification of organelles is the origin/speciation of eukaryotes

Currently, the relationship between the origin/diversification of organelles and the origin/speciation of eukaryotes remains controversial owing to the assumption of separate origin of organelles. While de novo organelle biogenesis in cyanobacterium TDX16 [77] demonstrates that nucleus, mitochondrion and chloroplast are formed all at once rather than separately, as such the origin/diversification of organelles surely leads to the origin/speciation of eukaryotes and vice versa. Hence the origin of organelles and the origin of eukaryotes as well as the diversification of organelles and the speciation of eukaryotes are two sides of the same coin and thus unified.

### 4.3 The origin/speciation of eukaryotes is the origin/speciation of unicellular eukaryotes

Eukaryotes comprise unicellular eukaryotes (protists) and multicellular eukaryotes, while the latter derive from the former. So, the origin of eukaryotes is undoubtedly the origin of unicellular eukaryotes. While the speciation of eukaryotes is usually considered to be the speciation of multicellular eukaryotes (e.g., animals and plants) that are easy to observe owing to their visible morphological and structural traits. This view is, however, incorrect. In truth, the speciation of eukaryotes is the speciation of unicellular eukaryotes (the progenitors) that determines the speciation of multicellular eukaryotes (the descendents). That is to say, the speciation of multicellular eukaryotes depends on and is achieved by the speciation of unicellular eukaryotes or new species of multicellular eukaryotes arise from unicellular rather than multicellular eukaryotes (multicellular eukaryotes can not give rise to new species of multicellular eukaryotes), while the putative hybrid speciation of multicellular eukaryotes that inter-specific sexual hybridization between two different species of plants or animals results in new species of plants or animals [80-84] is unachievable, because different species are reproductively isolated [85], and even if occasionally inter-specific hybridization occurs, the hybrid is, however, sterile or inviable (e.g., a mule is the infertile hybrid of a horse and a donkey). Virtually, except the rare case of hybrid-infertile/inviable inter-specific hybridization, all the observed and inferred hybridization events are, in essence, intra-specific hybridization between different subspecies, varieties, formae and ecotypes of the same species, lead only to new infraspecific (intra-specific) taxa, e.g., new subspecies, subvarieties, subformae and new ecotypes, but not new species, reproductively isolated from the parent species. Such that, the origin and speciation of eukaryotes can only be observed microscopically and are extremely difficult to capture. So, it is not surprise that organelle biogenesis in cyanobacterium TDX16 [77] is the only case of eukaryote speciation observed so far.

### 4.4 The origin and diversification of organelles share the same mechanism

Organelle biogenesis in the cyanobacterium TDX16 [77] is, as described above, a cellular event of organelle diversification/eukaryote speciation, showing that a cyanobacterium (prokaryote) turns into a new species of alga (eukaryote) by de novo biogenesis of new organelles after endosymbiotic acquisition of a preexisting alga's genome and hybridization of the acquired genome with its own one. Such that, the diversification of organelles for the speciation of eukaryotes, like the origin of organelles for the origin of eukaryotes, is also achieved by de novo biogenesis of organelles in prokaryotes, suggesting that the origin and diversification of organelles share the same cellular and molecular mechanism.

### 4.5 Genome hybridization hypothesis

In light of organelle biogenesis in the cyanobacerium TDX16 in combination with cellular and molecular evidence, we propose genome hybridization hypothesis (GHH) that the origin and diversification of organelles as well as the origin and speciation of eukaryotes are unified and achieved by genome hybridization: the endosymbiotic cyanobacteria/bacteria obtain the senescent/necrotic archaebacterial or eukaryotic hosts' genomes and hybridize with their own ones resulting in expanded genomes containing a mixture of hybrid prokaryotic genes and eukaryote specific genes, and thus the cyanobacteria/bacteria have to compartmentalize to accommodate different genes for specialized function of oxygenic photosynthesis (chloroplast), aerobic respiration (mitochondrion) and DNA preservation (nucleus), and consequently turn into photosynthetic/heterotrophic unicellular eukaryotes.

Cyanobacteria and bacteria symbiotic with archaebacteria are involved in the origin of organelles and origin of eukaryotes (Table 1); while cyanobacteria and bacteria symbiotic with eukaryotes are responsible for the diversification of organelles and speciation of eukaryotes (Table 2). The origin and diversification of nucleus and mitochondrion take place in both bacteria and cyanobacteria, while the origin and diversification of chloroplast occur only in cyanobacteria. This is because the capacity for organelle biogenesis depends on the metabolic capability of genome-recipient cyanobacteria and bacteria, but not genome-donor archaebacteria and eukaryotes. Bacteria are capable of respiration in which only mitochondrion can be formed; while cyanobacteria are capable of both photosynthesis and respiration in which chloroplast and mitochondrion can be developed.

**Table 1 Symbioses involved in the origin of organelles and eukaryotes**

| | Endosymbionts | Hosts (aerobic) | Origin of organelles | Origin of eukaryotes |
|---|---|---|---|---|
| 1 | Cyanobacteria | Archaebacteria | Nucleus, mitochondrion, chloroplast | Photosynthetic eukaryotes |
| 2 | Bacteria | Archaebacteria | Nucleus, mitochondrion | Heterotrophic eukaryotes |

**Table 2 Symbioses involved in the diversification of organelles and speciation of eukaryotes**

| | Endosymbionts | Hosts (aerobic) | Diversification of organelles | Speciation of eukaryotes |
|---|---|---|---|---|
| 1 | Cyanobacteria | Photosynthetic eukaryotes | Nucleus, mitochondrion, chloroplast | Photosynthetic eukaryotes |
| 2 | Cyanobacteria | Heterotrophic eukaryotes | Nucleus, mitochondrion, chloroplast | Photosynthetic eukaryotes |
| 3 | Bacteria | Heterotrophic eukaryotes | Nucleus, mitochondrion | Heterotrophic eukaryotes |
| 4 | Bacteria | Photosynthetic eukaryotes | Nucleus, mitochondrion, abnormal compartments* | Heterotrophic eukaryotes |

* occasionally formed structures with irregular shape and function.

Archaebacteria and eukaryotes (hosts) involved are all aerobic, because the endosymbiotic cyanobacteria/bacteria are oxygen-generating/requiring prokaryotes and thus can not inhabit oxygen-sensing/free anaerobic hosts. The archaebacteria and eukaryotes shelter the endosymbiotic cyanobacteria/bacteria from harsh conditions, e.g., high or low temperature; and in turn the endosymbiotic cyanobacteria/bacteria facilitate hosts' metabolism (e.g., synthesis of metabolites or nitrogen-recycling) rather than confer hosts the ability of photosynthesis or respiration. The hosts impose efficient control over the endosymbiotic cyanobacteria/bacteria under normal conditions, but turn senescence and lose control of these endosymbionts under adverse conditions. Such that the endosymbiotic cyanobacteria/bacteria take up their senescent/necrotic host's total DNA (genome) and retain the obtained DNA in a membrane-bounded heterogenous globular body (HGB) like those of TDX16[77] so as to keep from degradation.

After liberation from the dead host cell, the endosymbiotic cyanobacteria/bacteria hybridize the obtained DNA that was released from HGB with their own one, resulting in an expanded genome comprising hybrid prokaryotic genes (e.g., hybrid 16S rRNA) and completely novel or modified eukaryote specific genes (e.g., 18S rRNA). Such that, the cyanobacteria/bacteria are confronted with two problems:
(1) the expanded genome, a mixture of prokaryotic and eukaryotic genes, can no longer be deposited simply in the nucleoid, and must be categorized and preserved separately.
(2) the prokaryotic systems of respiration and photosynthesis are energetically less efficient to operate the expanded genome (genome expression) for upgraded cell functions. The endosymbiotic bacteria perform respiration in cytoplasmic membranes, which have a low energy-generating capacity due to the limited respiratory membrane surface, and so a specialized compartment with large and adjustable surface area for respiration is needed. The endosymbiotic cyanobacteria perform both photosynthesis and respiration in thylakoid membranes [86], which are also incapable of producing enough energy, because the concurrence of photosynthesis and respiration in thylakoid membranes reduce their energy-generating capacity and efficiency, and so photosynthesis and respiration must be performed separately in different compartments.

To tackle these two problems, DNA fragments with hybrid prokaryotic genes encoding constituents specific for photosynthesis and respiration (e.g., photosynthetic and respiratory electron transport chains), and essential components of prokaryotic protein synthesis systems (e.g., 16SrRNA, tRNA and ribosomal proteins) are sorted out and enclosed separately with membrane for specialized function of photosynthesis (chloroplast) and respiration (mitochondrion); while all the rest of DNA fragments containing eukaryote specific genes and nonfunctional DNA sequences are totally encased in a membrane-bounded compartment for preservation (nucleus). As a result, the organelle-less cyanobacteria and bacteria suddenly turn into organelle-containing unicellular photosynthetic eukaryotes and heterotrophic eukaryotes respectively.

### 4.6 The way of organelle biogenesis in cyanobacteria and bacteria

The biogenesis of chloroplast, nucleus and mitochondrion in cyanobacteria that had acquired their archaebacterial/eukaryotic hosts' genomes for the origin/speciation of photosynthetic eukaryotes proceeds in the same way as organelle biogenesis in the cyanobacterium TDX16 [77]; while the biogenesis of nucleus and mitochondrion in bacteria that had obtained their archaebacterial/eukaryotic hosts' genomes for the origin/speciation of heterotrophic eukaryotes follows the hypothetical way of organelle biogenesis in bacteria for the formation of cancer cells [87].

## 5. Testable predictions

### 5.1 Multiple origins of eukaryotes and organelles

Photosynthetic eukaryotes and heterotrophic eukaryotes origin from cyanobacteria and bacteria respectively by de novo biogenesis of organelles after endosymbiotic acquisition and hybridization of archaebacteria's genomes. There is no reason to think that only one or a limited number of cyanobacteria, bacteria and archaebacteria are involved in the origin of organelles and eukaryotes. In fact, the simple prokaryotes have a natural tendency to upgrade to complex eukaryotes and theoretically all aerobic bacteria, archaebacteria and cyanobacteria can participate in the origin of organelles and eukaryotes. So, eukaryotes and their organelles are of multiple origins, which sets the fundamental basis for biodiversity.

### 5.2 The formation of cancer cells is the speciation of eukaryotes

It has been proposed that cancer cells are new species of protists (e.g., HeLa cell) arising from bacteria [87]: the endosymbiotic bacteria in normal somatic cells turn into primary cancer cells for cancer initiation after acquisition and hybridization of the normal somatic cell's genome and de novo biogenesis of organelles; likewise the endosymbiotic bacteria in primary cancer cells turn into secondary cancer cells for cancer progression (e.g., invasion and metastasis) after acquisition and hybridization of the primary cancer cell's genome and de novo biogenesis of organelles. In such a way, the endosymbiotic bacteria in lower grade cancer cells (e.g., secondary cancer cells) turn into more invasive higher grade cancer cells (e.g., tertiary cancer cells) by acquisition and hybridization of the lower grade cancer cell's genome (e.g., the secondary cancer cell's genome) and de novo biogenesis of organelles. The primary cancer cells (containing DNA sequences derived from the normal somatic cell and the endosymbiotic bacterium), the secondary cancer cells (containing DNA sequences derived from the primary cancer cell and the endosymbiotic bacterium) and the higher grade cancer cells e.g., the tertiary cancer cells (containing DNA sequences derived from the secondary cancer cell and the endosymbiotic bacterium) are different species of protists. Hence, the formation of cancer cells is the speciation of eukaryotes that is achieved by de novo biogenesis of organelles in different endosymbiotic bacteria after acquiring and hybridizing the normal somatic cell or existing cancer cell's genome, that is to say, cancer is a speciation disease. Such that, cancer biology, evolutionary biology and cell biology are unified, while cancer cells provide the experimental material not only for cancer research but also for studying de novo biogenesis of organelles in prokaryotes and speciation of eukaryotes.

### 5.3 Organellar genomes are patchworks of diverse organisms' DNA sequences

During the origin and speciation of eukaryotes, genomic DNAs of diverse organisms including archaebacteria, bacteria, cyanobacteria and eukaryotes are intermingled, fragmented and hybridized resulting in a mixture of DNAs, which is subsequently partitioned into different compartments. So, each organelle's genome is a patchwork of diverse organisms-derived DNA sequences (so is the eukaryote's whole genome) and each organism-derived DNA sequences exist in all organelles' genomes. For example, a photosynthetic eukaryote's three organellar genomes are all assembled with diverse organisms' DNA sequences and if a cyanobacterium's gene presents in the mitochondrial genome, then similar or different DNA sequences derived from this cyanobacterium also exist in the chloroplast genome and nuclear genome.

### 5.4 Single source of archaebacterial DNAs in each eukaryote

Archaebacteria involve only in the origin but not the speciation of eukaryotes, and so all archaebacterial DNAs in each eukaryotic species come definitely form a single archaebacterium, which is important and useful for identifying the archaebacterium, inferring the origin or speciation of the eukaryotic species and verifying the multiple origins of eukaryotes.

### 5.5 Multiple sources of eubacterial and eukaryotic DNAs in each eukaryote

During eukaryote speciation, genomic DNAs of one eubacterium (cyanobacterium or bacterium) and one eukaryote that contains other eubacteria and euakaryotes' DNAs are introduced into the new species. Hence, eubacterial and eukaryotic DNAs in each eukaryotic species come from multiple eubacteria and eukaryotes except the ancestral eukaryotes that are formed during eukaryote origin containing only one eubacterium and one archaebacterium-derived DNAs but no other eukaryote-derived DNAs.

### 5.6 Different cellular traits of heterotrophic and photosynthetic eukaryotes

Heterotrophic and photosynthetic eukaryotes are different in origin and speciation, and apart from their well-known difference in organelles, e.g., the former possess lysosomes but no chloroplasts and vacuoles while the latter have chloroplasts and vacuoles but no lysosomes, they also exhibit other different cellular traits:

(1) The cytoplasmic membrane comprises one unit membrane in heterotrophic eukaryotes, but two unit membranes (cytoplasmic envelope) in photosynthetic eukaryotes that are closely appressed in most cases and usually misinterpreted as one unit membrane or two layers of one unit membrane, e.g., the cytoplasmic envelope of *Chroococcidiorella tianjinensis* [78].

(2) The nucleus is enclosed with two unit membranes (one set of envelope) in heterotrophic eukaryotes but four unit membranes (two set of envelopes) in photosynthetic eukaryotes, e.g., *Chroococcidiorella tianjinensis* [78]. The two set of nuclear envelopes is commonly misidentified as two nuclear membranes because each envelope's two membranes are usually in close apposition.

(3) Mitochondria are assembled de novo in the nuclei of heterotrophic eukaryotes but the chloroplasts of photosynthetic eukaryotes.

### 5.7 The outer envelope of the four membrane-bound chloroplast is the extended outer nuclear envelope

The chloroplasts in some algae (e.g., brown algae, diatoms and cryptophytes) are surrounded by four unit membranes (two set of envelopes), and the origin of the outer envelope that is cell biologically termed the "chloroplast endoplasmic reticulum" for the presence of ribosomes on its outer surface remains unclear [88]. It is explained that the four membrane-bound chloroplasts are secondary chloroplasts originated from algae in secondary endosymbiosis with the two membranes of the outer envelope being derived from the plasma membrane of the endosymbiotic algae and the phagosomal membrane of the hosts [89]. This explanation is completely wrong as the endosymbiotic hypothesis is disproved by our discovery [77]. Indeed, the four membrane-bound chloroplasts, like other chloroplasts, have only two membranes (one set of envelope), while their outer envelope is the outer nuclear envelope, which extends around the chloroplast functioning probably as the endoplasmic reticulum, and yet, as mentioned above, is usually misidentified as one unit membrane since its two membranes are tightly appressed in most cases.

### 5.8The non-photosynthetic plastids are not plastids

The presence of chloroplast-related genes in unicellular heterotrophic eukaryotes, e.g., apicomplexan parasites [90-92] and heterotrophic algae [93-95] is ascribed to the putative non-photosynthetic plastids. These putative non-photosynthetic plastids are, however, not plastids that exist in three types: the photosynthetic chloroplasts and the non-photosynthetic leucoplasts and chromoplasts. As summarized in Table1 and 2, chloroplasts can not be developed in bacteria due to their incapacity of photosynthesis. So, if bacteria endosymbiotically acquire photosynthetic eukaryotes' genomes, the chloroplast-related genes might be lost or deposited after hybridization in abnormal compartments and or in nuclei or mitochondria if there is no abnormal compartments. For example, the putative genome-less non-photosynthetic plastid of the heterotrophic green alga *Polytomella* spp [96] does not exist, while the chloroplast-related genes were lost or deposited in mitochondria. Therefore, the so-called non-photosynthetic plastids are not plastids, but abnormal compartments formed in bacteria during eukaryote speciation and in essence the heterotrophic algae are not algae that arise from cyanobacteria but bacteria-derived non-photosynthetic protists.

### 5.9 The nucleomorph is not the vestigial nucleus

The DNA-containing compartment in the four membrane-bounded chloroplasts (the so-called secondary chloroplasts) of cryptophytes and chlorarachniophytes is explained to be the vestigial nucleus of the endosymbiotic alga that gave rise to the secondary chloroplast, termed nucleomorph [97]. This explanation is wrong since the endosymbiotic hypothesis is invalidated. As discussed above in section 5.7, the four membrane-bounded chloroplasts are indeed double membrane-bounded chloroplasts surrounded by the outer nuclear envelopes, while the nucleomorph is actually not present in but enclosed along with the chloroplast by the extended outer nuclear envelope. In light of GHH, chloroplasts are formed only in cyanobacteria but not bacteria during eukaryote speciation, and given that (1) the non-photosynthetic plastids in heterotrophic eukaryotes, as mentioned above, are not plastids but abnormal compartments for preserving the chloroplast-related genes formed in bacteria that had obtained photosynthetic eukaryotes' genomes, and (2) the nucleomorph genes are eukaryotic in nature [98-100] yet different from the nuclear ones, e.g., 18S rRNA gene [100], it is logical and reasonable that if a cyanobacterium endosymbiotically acquires a non-photosynthetic plastid-bearing heterotrophic eukaryote's genome and hybridizes with its own one for speciation, the hybridized non-photosynthetic plastid-derived genes are noncanonical eukaryotic type and so encased in a double membrane-bounded compartment-the so-called nucleomorph in the way noted above in section 3 for individual preservation and then enclosed along with the chloroplast using the outer nuclear envelope so as to function as an integrated whole. Thus, the nucleomorph is an auxiliary compartment of chloroplast but not the vestigial nucleus, which is why its genome contains primarily house-keeping genes and a handful of chloroplast-associated genes but no genes for metabolic functions [101-106]. And so, the nucleomorph genome is a mosaic of diverse DNA sequences coming from at least two species of cyanobacteria, one species of bacterium and two species of eukaryotes (a non-photosynthetic plastid-bearing heterotrophic eukaryote and a photosynthetic eukaryote) .

### 5.10The amitochondriate eukaryotes congenitally lack mitochondrion

The presence of mitochondrion-related genes in amitochondriate eukaryotes (anaerobic protists) is ascribed to the putatively lost mitochondria. In truth, these putative mitochondria do not exist for the similar reason as discussed above for non-photosynthetic plastids. Anaerobic bacteria are incapable of aerobic respiration in which mitochondria can not be formed. Such that the aberrant anaerobic bacteria that endosymbiotically obtain their aerobic eukaryotic hosts' genomes turn into amitochondriate anaerobic protists with the mitochondrion-related genes being lost [107]or deposited after hybridization in nuclei[108-110]and or in abnormal compartments, e.g., hydrogenosome [111]. Therefore, the amitochondriate eukaryotes congenitally lack mitochondrion, and the putative mitochondrion-derived organelles[112]and mitochondrial remnants [113-115] are not true.

### 5.11 Different content of heterotrophic and photosynthetic eukaryotes' genomes

Heterotrophic eukaryotes' genomes contain only bacterial and archaebacterial but no cyanobacterial DNAs except the non-photosynthetic plastid-bearing parasites [90-92] and heterotrophic algae [93-96]. By contrast, photosynthetic eukaryotes' genomes contain cyanobacterial and archaebacterial or additional bacterial DNAs if heterotrophic eukaryotes are involved in speciation. So, photosynthetic eukaryotes fall into two categories: category1 contains only cyanobacterial and archaebacterial DNAs, while category 2 contains cyanobacterial, archaebacterial and bacterial DNAs, e.g., algae bearing the so-called nucleomorph [97-106].

### 5.12The heterotrophic eukaryote's nucleus contains a mitochondrial genome

Mitochondria are assembled in the primitive nucleus during the origin and speciation of heterotrophic eukaryotes, and after origin or speciation an entire mitochondrial genome (intact or fragmented mitochondrial DNA) remains in the nuclear genome enabling de novo biogenesis of mitochondria in the nucleus under stress conditions. So, the presence of an entire mitochondrial genome in the nuclei is not confined to the known cases [43, 48-49], but a general phenomenon in heterotrophic eukaryotes.

### 5.13The photosynthetic eukaryote's chloroplast contains a mitochondrial genome

Different from the case of heterotrophic eukaryotes, mitochondria are assembled in the primitive chloroplast during the origin and speciation of photosynthetic eukaryotes. So, after origin or speciation, the chloroplast genome retains a fragmented copy of mitochondrial DNA so as to develop mitochondria from scratch in the chloroplast when necessary in adverse conditions. It is noteworthy that the chloroplast genomes are larger in algae [116-119] and mosses[120] but smaller in ferns[121-124] and seed plants[125-131] in size than the corresponding mitochondrial genomes, yet the chloroplast genomes always contain more genes than the corresponding mitochondrial genomes in all photosynthetic eukaryotes. The expansion of mitochondrial genome size in ferns and seed plants is due primarily to the increase of repetitive sequences. So, the basic mitochondrial genomes of ferns and seed plants (containing only one copy of each repetitive sequence) are smaller in size than the corresponding chloroplast genomes, and it is more accurate to say the chloroplast genome of algae and mosses contains a mitochondrial genome, while the chloroplast genome of ferns and seed plants contains a basic mitochondrial genome. These mitochondrial genomes and basic mitochondrial genomes are highly fragmented with the DNA fragments being embedded in the corresponding chloroplast genomes, and so only a few genes (e.g., tRNA genes) remain intact while the rest of genes (e.g., rRNA genes) are broken into pieces that are separated by other coding or non-coding sequences.

### 5.14 Mitochondrial proteomes reside in nuclei of heterotrophic eukaryotes but chloroplasts of photosynthetic eukaryotes

The reason for the presence of some mitochondrial proteins in nuclei of heterotrophic eukaryotes [60-61] but chloroplasts of photosynthetic eukaryotes [57-59]as described in section 2.5 is unclear. In fact, like the case of DNA, it is also because mitochondria can be formed de novo in nuclei of heterotrophic eukaryotes but chloroplasts of photosynthetic eukaryotes. So, not just some but actually all of the mitochondrial proteins, the mitochondrial proteomes, reside in nuclei of heterotrophic eukaryotes but chloroplasts of photosynthetic eukaryotes.

### 5.15 The inferred organelle-to-organelle gene transfers do not exist

The common DNA sequences/genes in different organelles as described above in section 2.4 is inferred as the results of organelle-to-organelle gene transfers during or after the transition of the endosymbiotic cyanobacterium and bacterium into the chloroplast and mitochondrion, termed endosymbiotic gene transfer[132], interorganellar/intercompartmental gene transfer [133-134] and intracellular gene transfer [135]. The inferred organelle-to-organelle gene transfers, however, do not exist, because chloroplasts and mitochondria are not derived from cyanobacteria and bacteria as evidenced by de novo organelle biogenesis in the cyanobacterium TDX16 [77]. Indeed, the common DNA sequences/genes in different organelles result from genome hybridization and allocation of the hybrid DNA for organelle biogenesis during the origin and speciation of eukaryotes. Hence, the common DNA sequences/genes exist in organelles from the start of their assembly, reflecting their relationships in biogenesis. For example, mitochondria are assembled in the primitive nuclei of heterotrophic eukaryotes but the primitive chloroplasts of photosynthetic eukaryotes, and so mitochondria share more common DNA sequences with nuclei in heterotrophic eukaryotes but chloroplasts in photosynthetic eukaryotes.

### 5.16The inferred prokaryote/eukaryote-to-eukaryote gene transfers do not exist

The presence of homologous DNA sequences/genes in a eukaryote and a prokaryote or in two different eukaryotes is inferred as the results of prokaryote-to-eukaryote or eukaryote-to-eukaryote horizontal/lateral gene transfers (HGT/LGT)[136]. The inferred HGT/LGT are, however, unattainable, because eukaryotes are naturally resistant to foreign DNA. In fact, a eukaryote's genome is shaped during its origin or speciation by hybridizing the genomes of two parental organisms (two prokaryotes or a prokaryote and a pre-existing eukaryote) and thus inevitably contains DNA sequences/genes homologous to those of its parental prokaryote and eukaryote. Therefore, the homologous DNA sequences/genes exist in a eukaryote from the start of its origin or speciation but are not transferred from other organisms after its origin or speciation, which provides clues for inferring the eukaryote's origin or speciation.

### 5.17 The inferred phylogenies of eukaryotes and organelles are false

Each eukaryotic species arises directly from a cyanobacterium or a bacterium by de novo biogenesis of organelles after endosymbiotic acquisition and hybridization of an archaebacterium or a pre-existing eukaryote's genome. So, each eukaryote has two parental organisms, one is a cyanobacterium or a bacterium (genome-recipient), and the other is an archaebacterium or a eukaryote (genome-donor), and is formed by genomically hybridizing two into one instead of splitting one into two, a process similar in mechanism to sexual hybridization and sexual reproduction of eukaryotes, and thus the origin of sex seems to start from the origin and speciation of eukaryotes. While, current phylogenetic analyses are based on the wrong fundamental assumption that all eukaryotes and organelles descend directly from other eukaryotes and other organelles by putative gene duplication and point mutation. Therefore, the currently inferred phylogenies of eukaryote and organelles are false and misleading regardless of what algorithms have been applied.

### 5.18 The prokaryote-eukaryote evolutionary relationship is a web but not a tree

As described above, each heterotrophic/photosynthetic eukaryote arises from a bacterium/cyanobacterium that endosymbiotically acquires an archaebacterium or a eukaryote' genome. Thus, all eukaryotes (heterotrophic eukaryotes and photosynthetic eukaryotes) are evolutionarily linked with prokaryotes (archaebacteria, bacteria and cyanobacteria) in a non-hierarchical manner, and the evolutionary relationship of prokaryotes and eukaryote represents a truly interconnected web (web of life) rather than a bifurcating tree (tree of life), which is different, in essence, from the putative LGT/HGT-based net (reticulated tree) [137] or web [138].

## 6. Conclusion

The discovery of de novo organelle biogenesis in the cyanobacterium TDX16 [77] invalidates the endosymbiotic hypothesis and prompts an evidence-based reconsideration of the true origin of organelles and its association with the origin of eukaryotes and relationship with the diversification of organelles and speciation of eukaryotes that are important but unclear. GHH is a paradigm shift in understanding these questions. According to GHH, the origin/diversification of organelles and the origin/speciation of eukaryotes are unified and achieved by de novo biogenesis of organelles in bacteria or cyanobacteria after endosymbiotic acquisition and hybridization of their archaebacterial/eukaryotic hosts' genomes, while the formation of cancer cells is the speciation of eukaryotes as cancer cells are new species of heterotrophic protists arising from bacteria [87].Thus, GHH unifies evolutionary biology, cancer biology and cell biology and can account for all the evidence contradicting the endosymbiotic hypothesis and be verified with the testable predictions.